# Beam-based Feedback Simulations for the NLC Linac[*]

L. Hendrickson[#], N. Phinney, P. Raimondi, T. Raubenheimer, A. Seryi, P. Tenenbaum
SLAC, P.O. Box 4349, Stanford CA 94309 (USA)

*Abstract*

Extensive beam-based feedback systems are planned as an integral part of the Next Linear Collider (NLC) control system. Wakefield effects are a significant influence on the feedback design, imposing both architectural and algorithmic constraints. Studies are in progress to assure the optimal selection of devices and to refine and confirm the algorithms for the system design. We show the results of initial simulations, along with evaluations of system response for various conditions of ground motion and other operational disturbances.

## 1 INTRODUCTION

The NLC design specifies a sequence of beam-based trajectory feedback systems along the main linacs. This feedback is intended to operate on the average properties of a pulse train of 95 bunches at a repetition rate of 120 Hz. The measurements are beam position monitor (BPM) readings for a single pulse which may be averaged over a number of bunches. The corrections are primarily applied with fast dipole magnets but other devices were also studied. The feedback is designed to damp low frequency trajectory errors up to about 5 Hz. The system is modeled after the generalized feedback developed at the SLC [1] with improvements to provide more optimal response for the NLC.

Simulations of the NLC linac feedback systems have been performed in a Matlab environment [2] using the LIAR program [3] for tracking and beam calculations. Beam tests have also been done on the SLAC linac to confirm the simulation results [4].

## 2 MULTI-CASCADE SIMULATIONS

In the main linacs of both the SLC and NLC design, a series of feedbacks are used to keep the beam trajectory well centered in the quadrupoles and structures. To avoid overcorrection and ringing in response to an incoming disturbance, these systems need to communicate beam information. In the SLC, a simple one-to-one "cascade" system between loops was implemented. Each feedback controlled a number of beam states which were typically fitted beam positions and angles at a single point. On each beam pulse, these states were recalculated and the information passed to the next downstream loop before a correction was implemented. The downstream feedback would then use transport matrices to calculate the corresponding local states and subtract these from the current measurement to determine the residual beam motion to be corrected. The beam transport from the adjacent upstream feedback loop was calculated adaptively from the natural beam jitter.

In the presence of wakefields, the beam transport is non-linear and depends on the location where the perturbation occurs and distance it propagates. This means that the simple SLC single-cascade system is inadequate. In the NLC multi-cascade scheme, each feedback receives information from all of the upstream loops. The beam transport is an overconstrained least squares fit matrix, which converts many upstream states to the fitted downstream location.

The NLC simulations were done with five feedback systems distributed along the linac. Figure 1 shows the response of the last of these loops to a step disturbance early in the linac. The raw state is the measured beam orbit in the linac which can be fixed perfectly in simulation, when accelerator or modeling imperfections are not included. This simulation simply shows that the multi-cascade algorithm is correct. The response seen on the BPMs of the last feedback loop is identical to the design response for a single feedback loop, even though five loops respond. The cascade-adjusted state is the residual beam motion to be corrected by this feedback. Intuitively one might expect that the downstream loops would have nothing to do. However with the non-linear beam transport, the downstream feedbacks must make small corrections to achieve optimal system response.

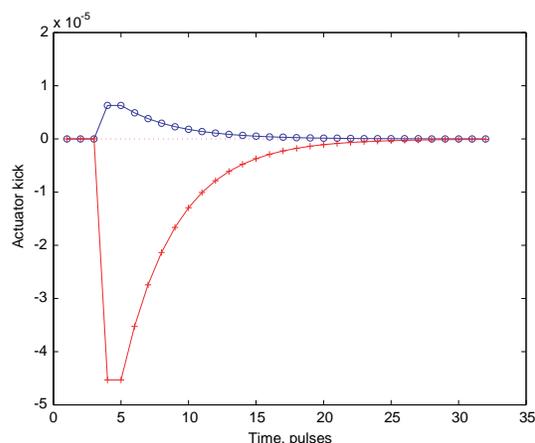

Figure 1: Multi-cascade simulation results. The lower plot (+) shows the real beam motion seen by the last NLC linac feedback following a perturbation. The upper plot (O) shows the cascade-adjusted state, which is the portion of the oscillation that this feedback corrects.

___
[*] Work supported by the U.S. Dept. of Energy under contract DE-AC03-76SF00515

[#] Email: ljh@slac.stanford.edu

## 3 DEVICE CONFIGURATION STUDIES

Another topic of study is the optimal number, type and placement of feedback devices for the NLC linac. In the SLC, the BPMs and correction dipoles were grouped together over a distance of at most 200 m because of bandwidth and connectivity constraints. The present NLC configuration includes 5 feedback systems distributed along the linac. Each feedback has 2 sets of 4 dipole correctors (two vertical and two horizontal). The first set of correctors is at the beginning of the section and the second set is located midway to the next feedback. Four sets of BPM measurements are used per loop, with 4 monitors in each set. Figures 2, 3 and 4 show the response to a step disturbance in the middle of the first loop with this configuration.

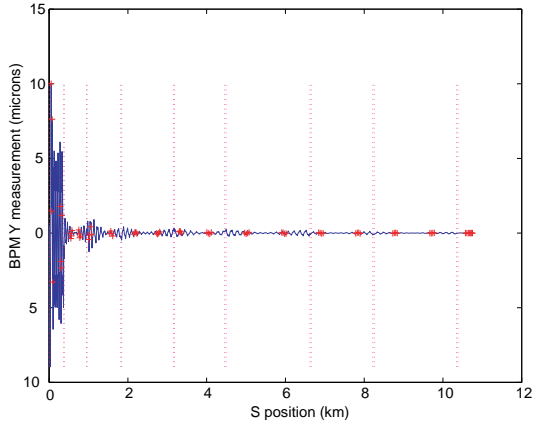

Figure 2: BPM orbit along the linac after feedback responds to a step function. The feedback correctors are located at the position of the vertical lines and the BPMs are marked with (+).

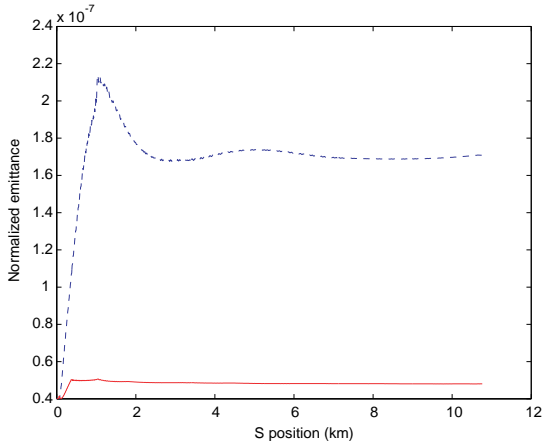

Figure 3: Simulation of emittance growth in the linac following a simple step disturbance, with (solid) and without (dashed) feedback.

When a bunch passes off-axis through the accelerating structures, the wakefields from the head of the bunch deflect the particles which follow causing a "tilt" or y-z correlation along the bunch. Local feedback corrects the bunch centroid but does not remove the tilt. The additional sets of BPMs and correctors help find a solution which minimizes both centroid offset and tilt. Figure 4 shows the offset of individual slices along the length of the bunch for the same simulation. The beam profile remains fairly flat for this device configuration.

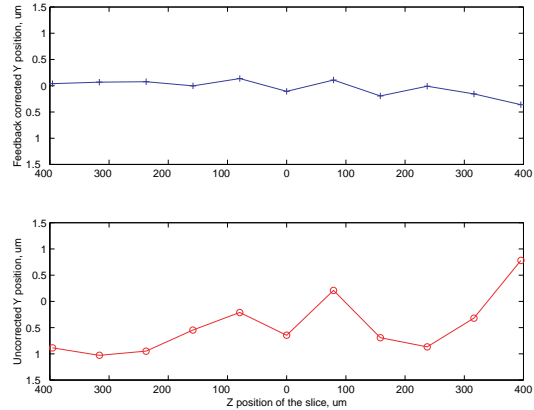

Figure 4: Offset of slices along the bunch at the end of the linac. The top figure shows the profile with NLC feedback on (+). The lower figure (O) shows the uncorrected bunch profile.

Most of the simulations have been done using dipole magnets as the correction devices. Other devices such as crab cavities or structure movers have also been studied to determine if they better correct the distortion of the bunches and the emittance dilution. Figure 5 shows the response to a perturbation for a system in which structure movers are used for one set of correctors in each feedback. The effectiveness of various alternatives is still under study.

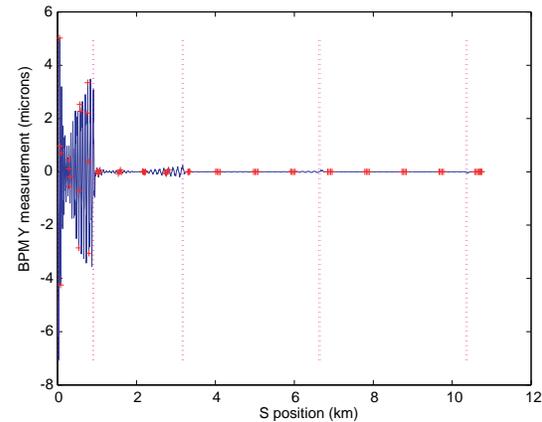

Figure 5: BPM orbit in the linac after a step response. Both dipole correctors and structure movers are used in this feedback configuration.

## 4 GROUND MOTION STUDIES

The critical test for a feedback system is performance in the presence of ground motion and accelerator errors. Simulations have been done using both the ATL ground motion model of Shiltsev [6] and the more complete model developed by Seryi [7]. Initial studies evaluated the RMS beam jitter and emittance growth after several seconds of ground motion changes, with feedback at 120 Hz rate.

The ATL ground motion model was used to compare the performance of different feedback configurations. The simulations used 30 minutes of ATL-like ground motion with a coefficient of 5.0e-7 $\mu m^2$/m/sec, a typical value for the SLAC site. The BPM resolution was 0.1 $\mu m$ and results from 100 random seeds were averaged. For the proposed NLC configuration, the emittance growth was less than 6%. Table 1 lists the results showing significantly larger emittance growth if less correctors or BPMs are used.

| # Feedback Loops | # BPMs per loop | # Cors per loop per plane | Emittance Growth (%) |
|---|---|---|---|
| 0 (off) | 0 | 0 | 104 |
| 5 | 16 | 2 | 31 |
| 5 | 8 | 4 | 21 |
| 5 | 16 | 4 | 5.7 |

Table 1: Emittance Growth for different feedback device configurations. Additional correctors and BPMs are effective in reducing emittance dilution.

Figure 6 shows how the emittance grows along the linac with the NLC feedback configuration. The emittance increases at the location of the corrector dipoles which create dispersion. However, the resulting emittance growth is only 5.7% compared with 104% without feedback, as shown in figure 7.

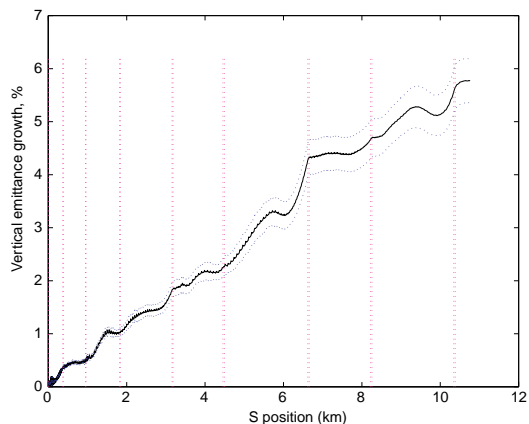

Figure 6: Simulation of vertical emittance along the NLC linac growth after 30 minutes of ATL ground motion, with feedback. Vertical lines mark the location of feedback corrector dipoles.

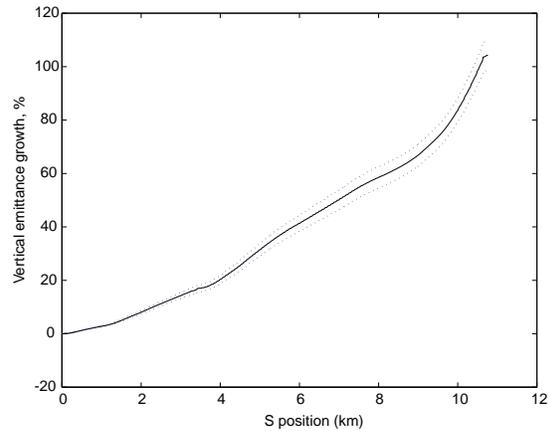

Figure 7: Simulation of vertical emittance growth after 30 minutes of ATL ground motion without feedback.

## 5 FUTURE PLANS

Further studies are needed to optimize the performance of the NLC feedback systems in the presence of ground motion and other errors. Integrated simulations with feedback, ground motion, cultural noise, and beam-based alignment algorithms are planned. These more complete simulations will be used to estimate the operational control requirements for the NLC and the beam jitter and emittance dilution expected during operation.

Another topic of study is the feasibility of adaptive or semi-invasive methods for measuring and updating the feedback model of beam transport in the linac. Because the planned feedback systems are distributed over large distances, accumulated energy or focusing errors can cause the actual model to differ significantly from the theoretical model. Present simulations assume that the model is updated by calibration measurements but adaptive methods would perform better.